\PassOptionsToPackage{unicode}{hyperref}
\PassOptionsToPackage{hyphens}{url}
\documentclass[preprint, nohyperref]{iacrtrans}
\usepackage{amsmath,amssymb}
\usepackage{lmodern}
\usepackage{iftex}
\ifPDFTeX
  \usepackage[T1]{fontenc}
  \usepackage[utf8]{inputenc}
  \usepackage{textcomp} 
\else 
  \usepackage{unicode-math}
  \defaultfontfeatures{Scale=MatchLowercase}
  \defaultfontfeatures[\rmfamily]{Ligatures=TeX,Scale=1}
\fi
\IfFileExists{upquote.sty}{\usepackage{upquote}}{}
\IfFileExists{microtype.sty}{
  \usepackage[]{microtype}
  \UseMicrotypeSet[protrusion]{basicmath} 
}{}
\usepackage{xcolor}
\usepackage{graphicx}
\makeatletter
\def\maxwidth{\ifdim\Gin@nat@width>\linewidth\linewidth\else\Gin@nat@width\fi}
\def\maxheight{\ifdim\Gin@nat@height>\textheight\textheight\else\Gin@nat@height\fi}
\makeatother
\setkeys{Gin}{width=\maxwidth,height=\maxheight,keepaspectratio}
\makeatletter
\def\fps@figure{htbp}
\makeatother
\newlength{\cslhangindent}
\newlength{\csllabelwidth}
\newlength{\cslentryspacingunit} 
 {
  \setlength{\parindent}{0pt}
  \ifodd #1
  \let\oldpar\par
  \def\par{\hangindent=\cslhangindent\oldpar}
  \fi
  \setlength{\parskip}{#2\cslentryspacingunit}
 }%
 {}
\usepackage{calc}

\usepackage{braket}

\newcommand{\dftwo}[1][n]{\mathcal{D}(\mathbb{F}_2^{#1})}

\ifLuaTeX
  \usepackage{selnolig}  
\fi
\IfFileExists{bookmark.sty}{\usepackage{bookmark}}{\usepackage{hyperref}}
\IfFileExists{xurl.sty}{\usepackage{xurl}}{} 
\usepackage{satex}

\newcommand{\thetitle}{The propagation game: on simulatability, correlation matrices, and probing security}
\newcommand{\theauthor}{Vittorio Zaccaria}
\institute{Department of Electronics, Information and Bioengineering\\Politecnico di Milano, Italy\\\email{vittorio.zaccaria@polimi.it}}

\newcommand{\finvect}{\textsf{FinVect}}
\newcommand{\finprob}{\textsf{FinProbVect}}
\newcommand{\finprobr}{\textsf{FinProbVect}$_\textsf{R}$}

\newcommand{\prop}{\textsf{PROP}}

\deftwocell[white]{xor: 2 -> 1}
\deftwocell[fill=gray]{and: 2 -> 1}
\deftwocell[black,fill=black]{dup: 1 -> 2}
\deftwocell[black,circle,fill=black]{erase: 1 -> 0}
\deftwocell[white,circle]{false: 0 -> 1}
\deftwocell[fill=black,circle]{random: 0 -> 1}
\deftwocell[fill=black,rectangle]{reg: 1 -> 1}
\deftwocell[fill=gray,circle]{true: 0 -> 1}
\deftwocell[crossing]{tau : 2 -> 2}

\hypersetup{
  pdftitle={\thetitle},
  pdfauthor={\theauthor},
  hidelinks,
  pdfcreator={LaTeX via pandoc}}

\title{\thetitle}
\author{\theauthor}
\keywords{side-channel attacks, Walsh transform, category theory, correlation matrices, string diagrams, prop categories}
\date{2022-12-03}

\begin{document}
\maketitle
\begin{abstract}
This work is intended for researchers in the field of side-channel attacks, countermeasure analysis, and probing security.
It reports on a formalization of simulatability in terms of categorical properties, which we think will provide a useful tool in the practitioner toolbox. The formalization allowed us to revisit some existing definitions (such as probe isolating non-interference) in a simpler way that corresponds to the propagation of \textit{erase morphisms} in the diagrammatic language of \prop{} categories. From a theoretical perspective, we shed light into probabilistic definitions of simulatability and matrix-based spectral approaches. This could mean, in practice, that potentially better tools can be built. Readers will find a different, and perhaps less contrived, definition of simulatability, which could enable new forms of reasoning. This work does not cover any practical implementation of the proposed tools, which is left for future work.

\end{abstract}

\section{Introduction}\label{introduction}

Masking is a common countermeasure to side channel attacks, which can pose a significant threat to hardware and software implementations of cryptographic primitives; however, it can be difficult to implement effectively, especially when considering more advanced adversary models such as probing adversaries or extended probing adversaries \cite{ishaiPrivateCircuitsSecuring2003, meyerConsolidatingSecurityNotions2019}.  

A gadget is considered secure $d$-probing if it is impossible to derive information about secret values encoded in masks or shares, even if an adversary has access to $d-1$ probes. 
The composability of two such gadgets, or the ability to determine whether their combination is also $d$-probing-secure, depends on the amount of refreshing randoms used to ensure non-interference \cite{bartheCompositionalVerificationHigherOrder2015}.

To assess the composability of gadgets, a technique called \textit{probing security by optimized composition} \cite{cassiersHardwarePrivateCircuits2020} is used, which takes advantage of the internal properties of the gadget to determine if the composition with other gadgets is secure. One such property is \textit{strong non-interference} (SNI)\cite{bartheStrongNonInterferenceTypeDirected2016}, which states that the number of input shares derivable from a set of probes depends only on the number of internal positions present in that set, as long as the size of the set is less than $d$. Establishing that a gadget is $d$-SNI may require lengthy proofs or the use of automatic tools \cite{bloemFormalVerificationMasked2018}, but once this property has been verified, the composition of the gadget can be studied using simpler rules, though not without difficulty.

This approach to gadget composability is called \textit{optimized} because it has the potential to lead to gadgets with minimal refresh efforts; however, it can be difficult to implement  \cite{moosGlitchResistantMaskingRevisited2019}. An alternative approach, called \textit{trivial composability}, aims to identify the inner properties of the gadget that make reasoning about composition even more straightforward, such that certain gadgets must ensure at least probe isolation-non-interference (PINI) to compose them \cite{cassiersHardwarePrivateCircuits2020}.

All forms of security proof are typically based on \textit{non-interference} \cite{bartheCompositionalVerificationHigherOrder2015}, that is, showing that some outputs of a system are not influenced by sensitive inputs. This is typically done by building probabilistic proofs of \textit{simulatability}, i.e. showing that the distribution of considered outputs is equal to the one of a system that does not depend on the sensitive values. 

In this work, we show that simulatability can afford a simple yet effective explanation through correlation matrices.
The formalization, based on category theory, allows for a simpler reexamination of existing definitions, such as PINI, in terms of repeated application of propagation rules to objects called \textit{erase morphisms}. From a theoretical perspective, this work provides a connection between probabilistic definitions of simulatability and matrix-based spectral approaches, which could potentially lead to the development of better tools and enable new forms of reasoning.

We will introduce background notions on correlation matrices and bit-vector distributions in Section \ref{background} while in Section \ref{on-the-algebra-of-simulatability} we will introduce the main theoretical result, i.e., a categorical/diagrammatic treatment of simulatability that stems from correlation matrices. As these are preliminary results, we conclude with a call for action for all the interested researchers, to collaborate on these new tools.

\section{Background}\label{background}
The theory of Boolean functions is important for side-channel attack analysis because side-channel attacks exploit physical characteristics of a cryptographic implementation, rather than attempting to directly break the cryptographic algorithm itself. These physical characteristics, known as side channels, can include information such as power consumption, electromagnetic radiation, or the execution time of the implementation.

Correlation immunity \cite{xiaoSpectralCharacterizationCorrelationimmune1988} is a property of Boolean functions that is related to side-channel analysis in that it can be used to design cryptographic implementations that are resistant to certain types of side-channel attacks \cite{zaccariaSpectralFeaturesHigherOrder2018,molteniSpectralFeaturesRobust2020}. Specifically, correlation immunity refers to the inability of an attacker to infer the value of a secret variable by measuring the correlations between the output of a Boolean function and the values of the secret variable.

In the context of cryptographic implementations, correlation immunity can be used to protect against side-channel attacks that aim to reconstruct secret information, such as a secret key, by analyzing the correlations between the output of the implementation and the values of the secret information. For example, an attacker may attempt to measure the power consumption of a cryptographic implementation while it is executing a Boolean function and observe that the power consumption varies depending on the values of the secret key. If the Boolean function is highly correlated with the secret key, then the attacker may be able to use this information to reconstruct the secret key.

The Fourier expansion of a Boolean function is a way of expressing the function as a linear combination of "base" functions, known as Fourier basis functions or parities. Each Fourier basis function corresponds to a certain subset of the input variables of the Boolean function, and the coefficients of expansion represent the contributions of each basis function to the overall output of the Boolean function.

The Fourier expansion of a Boolean function can be used to measure the correlation immunity of the function. This is because the magnitude of the expansion coefficients is indicative of the correlation between the function and the input variables. If the magnitude of a coefficient is large, then the corresponding input variable is strongly correlated with the output of the function. Conversely, if the magnitude of the coefficient is small, then the input variable is less correlated with the output of the function.

According to \cite{odonnelAnalysisBooleanFunctions}, one can define the Fourier expansion of \(f: \mathbb{F}_2^{n} \rightarrow \mathbb{R}\) by introducing an inner product
\[\braket{f|g} = 2^{-n} \sum_x f(x)g(x)\] and, under this product, one can define an orthonormal basis set composed of functions called \emph{parities}: \(\chi_\gamma(x) = (-1)^{{{\gamma}^\intercal} x}\); any pseudo-Boolean function $g$ can be represented as a linear combination of these parities:
\[g(x) = 2^{-n}\sum_\gamma T_g(\gamma)(-1)^{{{\gamma}^\intercal}x}\]
where the term
\[T_{g}(\gamma)= \sum_{x \in \mathbb{F}_2^{n}} g(x)(-1)^{{{\gamma}^\intercal} x}\]
is called the Fourier expansion of \(g(x)\). Borrowing the notation from quantum mechanics (for reasons that will become evident in the following pages), we will use the symbol \(\ket{\gamma} = \chi_\gamma(x) = (-1)^{{{\gamma}^\intercal} x}\) and
write \[g(x) \propto \sum_\gamma T_g(\gamma)\ket{\gamma}\] A remarkable property of the Fourier expansion is that the study of any Boolean function \(g(x)\) can be
reduced to the study of vectors \(T_g \in \mathbb{R}^{2^n}\). Correlation immunity can, in fact, be observed directly through the Fourier coefficients. For example, assume that \(g(x)\) is decomposed into the following combination of parity functions:
\[g(x) \propto a\ket{00} + b \ket{01} + c\ket{10} + d\ket{11}\] then,
\[T_{g}(\gamma) = {{\begin{bmatrix} a & b & c & d\end{bmatrix}}^\intercal}\]represents the correlations of \(g\) with each of the orthonormal bases, and the inner product between functions \(\braket{f|g}\) can be understood as the dot product of vectors $T_g$ and $T_f$. In particular, the linear form \(\braket{f|-} = \bra{f}\) is representable in \(\mathbb{R}^{2^n}\) with a row vector \({{T_f(\gamma)}^\intercal}\); we'll call this linear form a \emph{co-vector}.

\subsection{The Walsh transform}\label{the-walsh-transform}
The Walsh transform is an extension of the Fourier expansion above to any vectorial Boolean function
\(f: \mathbb{F}_2^{n}\rightarrow \mathbb{F}_2^{m}\) in the space \(\mathbb{R}^{2^m} \times \mathbb{R}^{2^n}\). The Walsh transform is defined as a \(2^{m}\times2^{n}\) matrix \(\widehat{f}\) whose elements
are:
\[\widehat{f}_{\omega,\alpha} = \sum_{x \in \mathbb{F}_2^{n}} (-1)^{{{\omega}^\intercal} f(x) \oplus {{\alpha}^\intercal} x}\]
Sometimes, they appear in the literature scaled by a coefficient \(2^{-n}\), and thus called \emph{correlation matrices} \cite{daemenCorrelationMatrices1995}:
\[W_{f} =2^{-n} \widehat{f}\] Each row of the correlation matrix is in fact a Fourier expansion of a Boolean function \({{\omega}^\intercal}f(x), \omega \in \mathbb{F}_2^{m}\). Some notable properties are \cite{parriauxSpectralApproachDesign2011}:
\begin{itemize}
\item  If \(W_f\) is orthogonal (i.e.~\(W_f {{W_f}^\intercal} = I\)) then it is also \textit{balanced} \cite{daemenCorrelationMatrices1995}.
\item  Any orthogonal \(W_f\) is also invertible and vice-versa.
\end{itemize}

An important case is when \(f(x) = Mx\) (where \(M\) is an invertible matrix) \cite{zaccariaSpectralFeaturesHigherOrder2018}. This case can describe a circuit that, for example, decodes a set of shares into the corresponding secret value or vice-versa: \[W_f = [\delta_{i, {{(M^{-1})}^\intercal}j}]_{i,j}\]

\subsection{Bit-vector probability distributions}\label{bit-distributions}
Bit-distributions are probability distributions over bit-vectors, or sequences of $m$ bits: $$\dftwo[m] = \mathbb{R}^{\mathbb{F}^m_2} \simeq \mathbb{R}^{2^m}$$They are often used in cryptography to model random events or to sample from distributions of secret keys or random numbers. Bit distributions are thus among the most prominent examples of pseudo-Boolean functions. 

They are important for another reason; they can be used to reason about non-interference and simulatability.
We start by analysing 1-bit distributions and  assume that \(x\) is a bit with a non-uniform probability \(a\)
of being \(1\) \[p_X(x) = \delta_{x,0}(1-a) + \delta_{x,1}a\] then
\[T_{p_X}(\gamma) = \sum_{x \in \mathbb{F}_2^{n}} p_X(x)(-1)^{{{\gamma}^\intercal} x}= \sum_{x \in \mathbb{F}_2^{n}} [\delta_{x,0}(1-a) + \delta_{x,1}a](-1)^{{{\gamma}^\intercal} x} = (1-a) + a(-1)^\gamma\]
In terms of bracket notation: \[p_X(x) = \ket{0} + (1-2a)\ket{1}\]For
uniformly random values we have (\(a = 1/2\)) this becomes
\(p_X(x) = \ket{0}\), i.e.,

\begin{equation}
        T_{p_X}(\gamma) = \begin{bmatrix} 1 \\ 0 \end{bmatrix} 
        \label{eq:unif}
\end{equation}

When \(a\)
is either 0 or 1, we deal with a fixed value \(x=s\) where \(s \in \{F, T\}\); its probability will be
\[p_X(x) = \ket{0} + (-1)^s\ket{1}\]

This reasoning can be extended to the distribution of multiple bits through the joint probability distribution of two independent Boolean variables $X$ and $Y$:
\[p_{X,Y}(x,y) = p_X(x)p_Y(y)\] By expanding both distributions in terms of parities, we get:
\begin{equation}
p_{X,Y}(x,y)= \sum_{\gamma}\sum_{\zeta}T_{P_X}(\gamma)T_{P_Y}(\zeta)\ket{\gamma \zeta}
\label{eq:joint}
\end{equation}
where we denote \(\ket{\gamma \zeta}\) the orthonormal bases of the joint space of functions over \(X \otimes Y\).

\subsubsection{Example}\label{example}

Assume \(n=3\) variables \(Q=(X,R_1,R_2)\) where \(X=s\) while \(R_*\)
are independent uniform randoms; we have that
\[ {p_{Q}}(x,r_1,r_2) = (\ket{0} + (-1)^s\ket{1})  \ket{0} \ket{0}=\ket{000} + (-1)^s\ket{100} \]
which gives a direct representation for \(T_{Q}\) i.e.:
\[\begin{bmatrix} 1 \\ 0 \\ 0 \\ 0 \\ (-1)^s \\ 0 \\ 0 \\ 0 \end{bmatrix}\]

\subsection{The Walsh transform as a map}\label{the-walsh-transform-as-a-map}
Walsh transforms and probability distributions combine to describe the probabilistic behavior of a circuit \cite{daemenCorrelationMatrices1995,parriauxSpectralApproachDesign2011}; in fact, given a function
\[ y = f(x), ~f: \mathbb{F}_2^{n}\rightarrow \mathbb{F}_2^{m}\]and a probability distribution \(p_X: \dftwo[n]\) (which maps from each combination of bit values to its probability), the following relation holds: \[W_{f}T_{p_X} = T_{p_Y}\]where \(p_Y: \dftwo[m]\) is the distribution of the output value \(y\). For example, assume \(f(x)=[f_0(x), f_1(x), f_2(x)]\), where

\begin{align}
        f_0 & = x_0 \oplus x_1 \oplus x_2 \\
        f_1 & = x_1 \\
        f_2 & = x_2 \,
\end{align}

and that $x = (s, r_1, r_2)$ where $s$ is a constant value, while $r_1, r_2$ are two uniformly random values. Indeed, we have that the distribution of output shares is the following:

\begin{equation}
T_{p_Y}(\gamma)  = W_{f} T_{p_X}(\gamma) = \begin{bmatrix}
1 & 0 & 0 & 0 & 0 & 0 & 0 & 0 \\
0 & 1 & 0 & 0 & 0 & 0 & 0 & 0 \\
0 & 0 & 1 & 0 & 0 & 0 & 0 & 0 \\
0 & 0 & 0 & 1 & 0 & 0 & 0 & 0 \\
0 & 0 & 0 & 0 & 0 & 0 & 0 & 1 \\
0 & 0 & 0 & 0 & 0 & 0 & 1 & 0 \\
0 & 0 & 0 & 0 & 0 & 1 & 0 & 0 \\
0 & 0 & 0 & 0 & 1 & 0 & 0 & 0 \\
\end{bmatrix}\begin{bmatrix} 1 \\ 0 \\ 0 \\ 0 \\ (-1)^s \\ 0 \\ 0 \\ 0 \end{bmatrix} = \begin{bmatrix} 1 \\ 0 \\ 0 \\ 0 \\ 0 \\ 0 \\ 0 \\ (-1)^s  \end{bmatrix}
\label{eq:example1}
\end{equation}
In fact, this encodes $s$ on the three shares\footnote{Note that the three shares are not decomposable anymore in three independent   variables. Somehow these become \emph{entangled} to highlight yet again the similarity with quantum computing.}:
\[p_Y(y) = (\ket{000} + (-1)^s\ket{111}) \]

\section{On the algebra of simulatability}
\label{on-the-algebra-of-simulatability}

Let us now depart from linear algebra into a small category-theoretic tour, which starts by considering that matrices are linear maps over vector spaces. Indeed, we consider 
the category of finite-dimensional vector spaces (\finvect) that has vector spaces as objects and matrices as morphisms,
the latter respecting obvious composition properties. 
\finvect{} is also a symmetric \textit{monoidal} category, in the sense that one can define, for all objects and morphisms, a tensor product that abides by well-known \textit{pentagon} rules \cite{maclaneCategoriesWorkingMatematician1969}. For \finvect{} this is the actual tensor product of vector spaces. 

Now, each $\dftwo[n]$ (for any $n$) is evidently a vector space and correlation matrices \textit{map} to vector spaces of the same type. 
In fact, we can define the category with objects $\dftwo[n]$ and correlation matrices as morphisms as a subcategory of \finvect  $$\finprob \subseteq \finvect$$

More interestingly, recalling the joint probability distributions (Eq. \ref{eq:joint}), each of these vector spaces can be seen as a tensor product of a single generating vector space. 
\begin{equation}
        \dftwo[n] = \bigotimes_n \dftwo[1] 
\label{eq:prop1}
\end{equation}

It turns out that property in Eq. \ref{eq:prop1} allows us to reason about \finprob{} in simpler terms. In fact, if we work in the context of objects of the type $\dftwo[*]$ we could say that a correlation matrix mapping  $\dftwo[n] \to \dftwo[m]$ is just a morphism $n \to m$, while the tensor product $M_0 \otimes M_1$ of matrices $M_0: \dftwo[n_0] \to \dftwo[m_0]$ and $M_1: \dftwo[n_1] \to \dftwo[m_1]$ is a morphism $n_0 + n_1 \to m_0 + m_1$. This type of description is associated with a product and permutation (\prop{}) category \cite{fongSevenSketchesCompositionality2018}.

A \prop{} is a strict symmetric monoidal category with natural numbers as objects, where the monoidal tensor $\otimes$ works on objects as the sum over natural numbers. 
In a \prop{}, only connectivity matters \cite{coeke2017}; Transformations between functions depend on the morphisms themselves and not on the objects to which they apply. 

Any \prop{}s can be summarized by a "presentation" $(\Sigma, E)$ that corresponds to a set of generator morphisms $\Sigma$ and a set $E$ of equations (or equivalences) between morphisms. 
\prop{}s can be amenable to simple syntactic reasoning in the form of diagrams. In particular, \prop{} \finprob{} is a 'semantic' category which can be further abstracted into a 'syntactic' one where morphisms are represented as \textit{string} (or \textit{tensor}) diagrams or \textit{signal flow graphs} \cite{fongSevenSketchesCompositionality2018}. Reasoning with such diagrams is \textit{sound}, that is, two diagrams represent the same correlation matrix if one can be turned into the other, and \textit{complete} i.e., if two diagrams represent the same matrix, then one can be turned into the other.

\subsection{The \finprob{} \prop{} category} It is well known that all Boolean
functions can be represented through \textit{algebraic normal form} (ANF, \cite{carletBooleanFunctionsCryptography}).
This means that any Boolean function can be built using two operators
($\oplus$ and $\wedge$) and a constant (1 or \textit{true}). 
This suggests a minimum set of generating morphisms/correlation matrices $\Sigma^*$ in the \finprob{} category:
$$\Sigma^* = \{ W_{\oplus}, W_{\wedge}, W_1 \} $$
This set is, however, not enough for \finprob{} as one must explicitly introduce copy and erasure of variables $$\Delta: (x) \mapsto (x,x), !: (x) \mapsto () $$ which is somehow implicit in the expression language (but will be explicit in the diagrammatic one); additionally, it will be useful to introduce also a morphism for the 0 (false) value:
$$\Sigma = \{ W_{\oplus}, W_{\wedge}, W_1, W_0, W_{\Delta}, W_{!}\} $$
where, a part from correlation matrices of $\oplus$ and $\wedge$ operators, $$W_0 = \begin{bmatrix} 1 \\ 1 \end{bmatrix}, W_1 = \begin{bmatrix} 1 \\ -1 \end{bmatrix}, W_\Delta = \begin{bmatrix} 1 & 0 \\ 0 & 1 \\ 0 & 1 \\ 1 & 0  \end{bmatrix}, W_!=\begin{bmatrix} 1 & 0 \end{bmatrix} = \bra{0}$$ 

Syntactically, we can represent the above matrices as the following symbols in the \textit{signal flow graph} category:

\[ 
        \twocell{xor} \:=\: W_{\oplus} \qquad  
        \twocell{and} \:=\: W_{\wedge} \qquad
        \twocell{false} \:=\: W_0 \qquad 
        \twocell{true} \:=\: W_1 \qquad 
\]

\[
        \twocell{dup} \:=\: W_{\Delta} \qquad
        \twocell{erase} \:=\: W_! 
\]

\begin{example}
For example, let us consider the example function  $f$ in Eq. \ref{eq:example1}; this can be translated
into a transformation of probabilities $$W_f: \bigotimes_3 \dftwo[1] \to \bigotimes_3 \dftwo[1]$$ as:
\[
        W_f = (W_{\oplus} \otimes I \otimes I)(I \otimes \sigma \otimes I)(W_\oplus \otimes I \otimes W_\Delta)(I \otimes W_\Delta \otimes I)
\]
Reading the equation right to left, we can produce a syntactic, signal flow graph view of the above transformation:
\[
        \twocell{(1 * dup * 1) * (xor * 1 * dup) * (1 * tau * 1) * (xor * 2)}  
\]
to be interpreted top to bottom.
\end{example}

\finprob{} is equipped with equations as well. These are mostly related to the monoidal nature of the $\oplus$ and $\wedge$ operators and we can use syntactic diagrams to represent them,
\[
        \twocell{(1 * false) * (xor)} = \twocell{(false * 1) * (xor)} = \twocell{(1) * (1)} \qquad \twocell{(1 * true) * (and)} = \twocell{(true * 1) * (and)} = \twocell{(1) * (1)}
\]

An additional equations are related to the duplication and erase morphisms:

\begin{equation}
\twocell{(dup) * (1 * erase)} = \twocell{(dup) * (erase * 1)} = \twocell{(1) * (1)}
\label{eq:erase-a}
\end{equation}

The latter corresponds to:

\[(W_{Id} \otimes \bra{0}) W_\Delta \simeq   (\bra{0} \otimes W_{Id}) W_\Delta \simeq W_{Id}\] where \(W_\Delta\)
and can be readily demonstrated
\[(\bra{0} \otimes W_{Id}) W_\Delta = (\begin{bmatrix} 1 & 0 \end{bmatrix} \otimes Id) \begin{bmatrix}1 & 0 \\ 0 & 1 \\0 & 1 \\ 1 & 0 \end{bmatrix} = \begin{bmatrix} Id & \bar{0} \end{bmatrix} \begin{bmatrix}1 & 0 \\ 0 & 1 \\0 & 1 \\ 1 & 0 \end{bmatrix} = Id = W_{Id}\]

Analogously:
\begin{equation}
        \twocell{(xor) * (erase)} =\twocell{(and) * (erase)} = \twocell{(1 * 1) * (erase * erase)}
\label{eq:erase-b}
\end{equation}

\subsection{The \finprobr{} \prop{} category}

The erase morphism ($!$) will play a role in the following for two reasons; on the one hand, it might be used in the situation where, from a vector Boolean function, we take a subset of outputs. The "propagation" of the erase morphism given by Eq. \ref{eq:erase-a} and Eq. \ref{eq:erase-b} can be used to derive the correlation matrix of such a reduced function. On the other hand, the dual morphism of \textit{erase} is $$!^{\dagger} = \ket{0} = \begin{bmatrix}1 \\ 0 \end{bmatrix} = \twocell{random}$$ which is just the Fourier transform of a uniform random value (Eq. \ref{eq:unif}); this is in fact the correlation matrix of a $0 \to 1$ gate that produces a uniform random.

We can think of a new \prop{} category called \finprobr{} where the signature is equipped with all the morphisms of $\finprob$ plus the random gate $\ket{0}$. This allows us to introduce a new syntactic equation called the "cut" rule:

\begin{equation}
        \twocell{(random * 1) * (xor)} = \twocell{(erase) * (random)}
\end{equation}

Proof:

\[W_\oplus(\ket{0} \otimes Id) \simeq \ket{0} \bra{0}\] readily provable
as: \[W_\oplus(\ket{0} \otimes Id) = \begin{bmatrix}1 & 0 & 0 & 0 
\\ 0 & 0 & 0 & 1 \end{bmatrix}(\begin{bmatrix}1 
\\ 0 \end{bmatrix} \otimes Id) = \begin{bmatrix}1 & 0 & 0 & 0 
\\ 0 & 0 & 0 & 1 \end{bmatrix}\begin{bmatrix}Id 
\\ \bar{0} \end{bmatrix}=\begin{bmatrix}1 & 0 \\ 0 & 0 
\end{bmatrix} = \ket{0} \bra{0}\] 

Note that the above equation is not true for any other gate; let us consider, for example, the $\wedge$ gate:

\begin{equation}
        \twocell{(random * 1) * (and)} \neq \twocell{(erase) * (random)}
\end{equation}

As one can verify through the resulting probability distribution: 
\[
W_\wedge(\ket{0} \otimes Id) = 
\begin{bmatrix} 1 & 0 & 0 & 0  \\ \frac{1}{2} & \frac{1}{2} & \frac{1}{2} & -\frac{1}{2} \\
\end{bmatrix} \begin{bmatrix}Id 
\\ \bar{0} \end{bmatrix}= \begin{bmatrix}1 & 0 \\ \frac{1}{2}  & \frac{1}{2} 
\end{bmatrix} 
 \]

\subsection{Simulatability as a "propagation" game} 
Simulatability is all about proving the equivalence of two output distributions up to a certain constraint\footnote{Here output is a generalised term which might refer to actual outputs or internal values (probes).}; in the context of $d$-probing security, the first one is the one of the original circuit $f$, while the second one is the same circuit as viewed by a $d$-limited adversary $f_d$, i.e., an adversary that has access only up to $d$ inputs of the circuit with the remaining $n_i-d+1$ inputs are taken as uniformly random:

$$p_f = p_{f_d}$$

If such an equivalence holds then $f_d$ is called the \textit{simulator}. Constructing the simulator means proving the above equivalence and, given the property of secret-sharing, no information can be derived from the actual encoded secret.
While having a constructive proof is useful, sometimes we want only to know if it exists, and not how it is built. 
\finprobr{} allows us to syntactically and soundly represent the existence of a simulator problem 
as a pure "rewriting" game according to the equations of the presentation itself.

\begin{figure}
\centering
\includegraphics{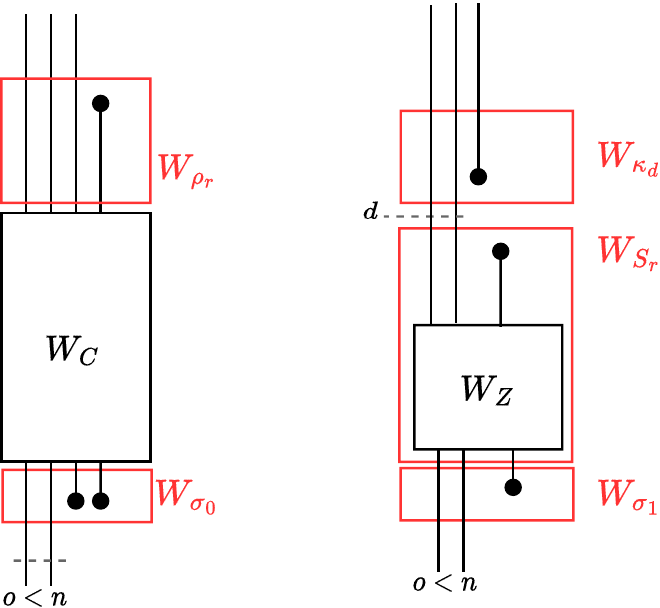}
\caption{\label{fig:sim}A simulator $S_r$ is just a randomized circuit with $d$ inputs that produces the same output distribution as $C$ as long as $o \leq n$.}
\end{figure}

Consider the correlation matrix of a circuit \(C\) with \(n\) outputs (See fig. \ref{fig:sim}, left). We say that \(o\) outputs
of a circuit \(C\) can be \textbf{simulated} with \(d\) inputs if and only if one can prove the equivalence with the circuit in fig. \ref{fig:sim}, right, i.e., if one diagram can be deduced from another by using the equation of the \finprobr{} presentation. 

More precisely, $C$ can be simulated if we can refactor
\(W_{\sigma_0}W_CW{\rho_r}\) (where \(\sigma_0\) is \(n-o\) erase morphisms 
and otherwise identities and \(\rho_r\) is a matrix with \(r\) randoms
and otherwise identities) into
\(W_{\sigma_1} W_{S_r} W_{\kappa_d}\) where \(\kappa_d\) is a matrix
of up to \(d\) identities and otherwise erase morphisms.

We call \(W_{S_r}\) the correlation matrix of the \textbf{simulator} of the
circuit (which is a probabilistic circuit, as it contains randoms). We have just defined simulatability not as much as identities (or probes) propagating backward (e.g., \cite{cassiersHardwarePrivateCircuits2020}) but as \textbf{erase morphisms propagating back according the \prop{} presentation}.

\begin{example}
The cut rule:

\begin{equation}
        \twocell{(random * 1) * (xor)} = \twocell{(erase) * (random)}
\end{equation}
embodies the smallest form of simulatability; i.e., if $$C = \twocell{xor}, \qquad W_{\rho_r} = \twocell{(random * 1)}$$
the output can be simulated with zero identities $$W_{\kappa_d} = \twocell{erase}$$ 
\end{example}

\begin{example}
As another example we show that, contrary to what has been
observed in \cite{bartheStrongNonInterferenceTypeDirected2016}, the \texttt{refreshM2} gadget does not depend on any input share at all for the considered outputs (see Fig. \ref{fig:refresh}).

\begin{figure}
\centering
\includegraphics[width=.7\textwidth]{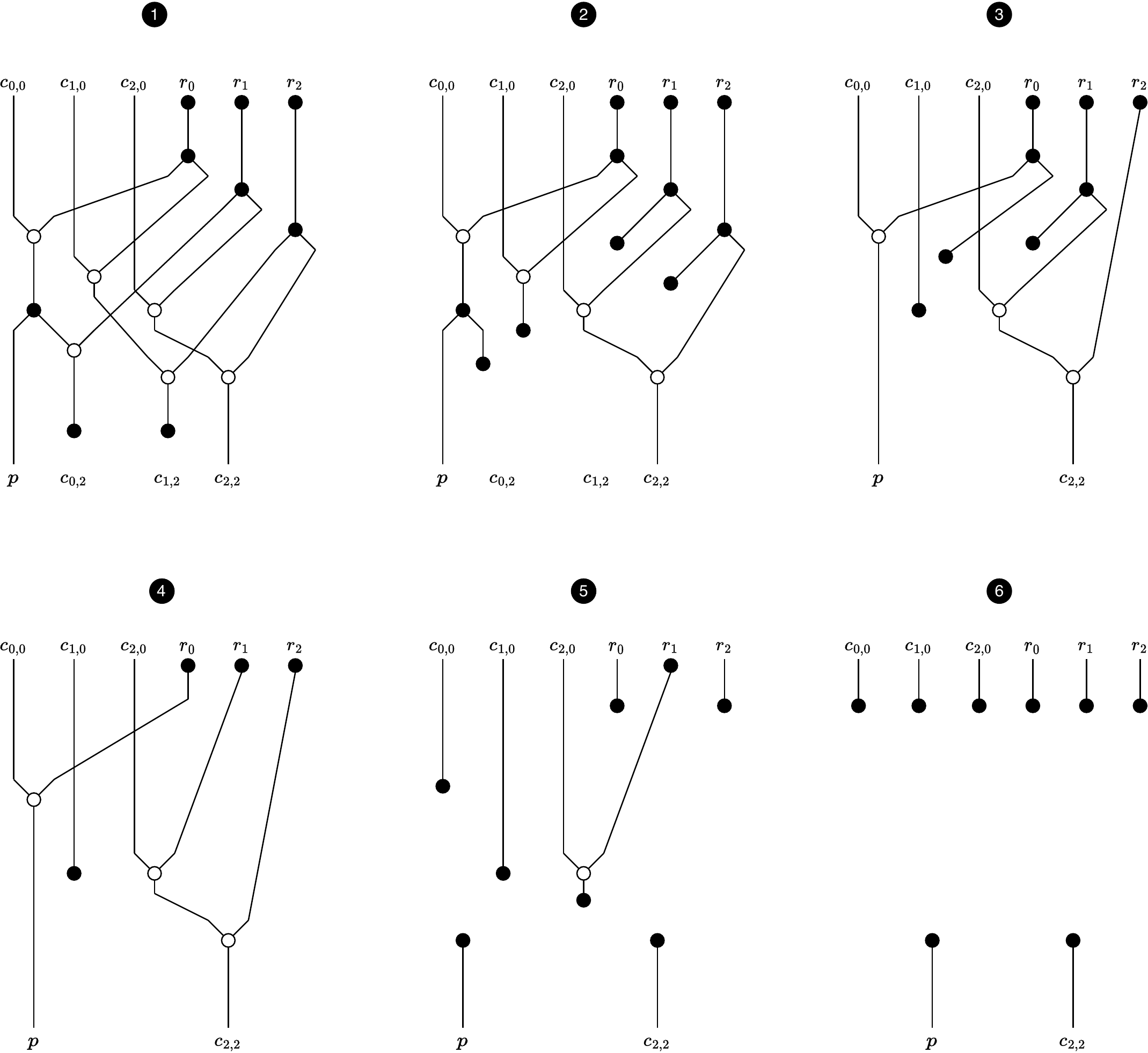}
\caption{By applying the \prop{} presentation equations, erase morphisms propagate back and allow the application of cut rules. In the end, both $p$ and $c_{2,2}$ are simulatable with zero inputs. \label{fig:refresh}}
\end{figure}
\end{example}

\subsection{Redefined simulatability can explain Probe Isolation Non
Interference}
We will prove that PINI (probe isolating non-interference, \cite{cassiersHardwarePrivateCircuits2020,cassiersHardwarePrivateCircuits2021}) is a composable property and appreciate how much erase propagation can simplify the reasoning.

We say that an input domain ``receives an erase morphism'' when, applying \finprobr{}'s equations, one can rewrite the circuit correlation matrix such that matrix \(W_{\kappa_d}\) presents an erase morphism on that particular input domain.

Define \(R_q(n)\) the predicate that asserts that \(n\) input domains to
circuit \(q()\) receive an erase morphism, \(p_q\) and \(o_q\) respectively
the probes in \(q\) and the output domains taken from \(q\).

The PINI property for \(q\) can be defined as a simple predicate:
\[p_q + o_q <t \implies R_q(t-(p_q + o_q))\] This property is
composable. Consider in fact the composition of two gadgets
\(f \circ g\), where for both \(f\) and \(g\)
\begin{align}
p_f + o_f <t \implies R_f(t-(p_f + o_f)) \\
p_g + o_g <t \implies R_g(t-(p_g + o_g)) \label{eq:snd}
\end{align}
When \(f\) is composed
with \(g\) (i.e., \(f \circ g\)), the outputs domains taken from \(g\) are the
ones which will not receive an erase, i.e.: \(o_g = p_f + o_f\). If
we put this definition in Eq. \ref{eq:snd} we get
\[p_g + p_f + o_f <t \implies R_{f\circ g}(t-(p_g + p_f + o_f))\] which
is the PINI property for \(f \circ g\). 

\subsubsection{Example}
\begin{figure}
\centering
\includegraphics[width=.5\textwidth]{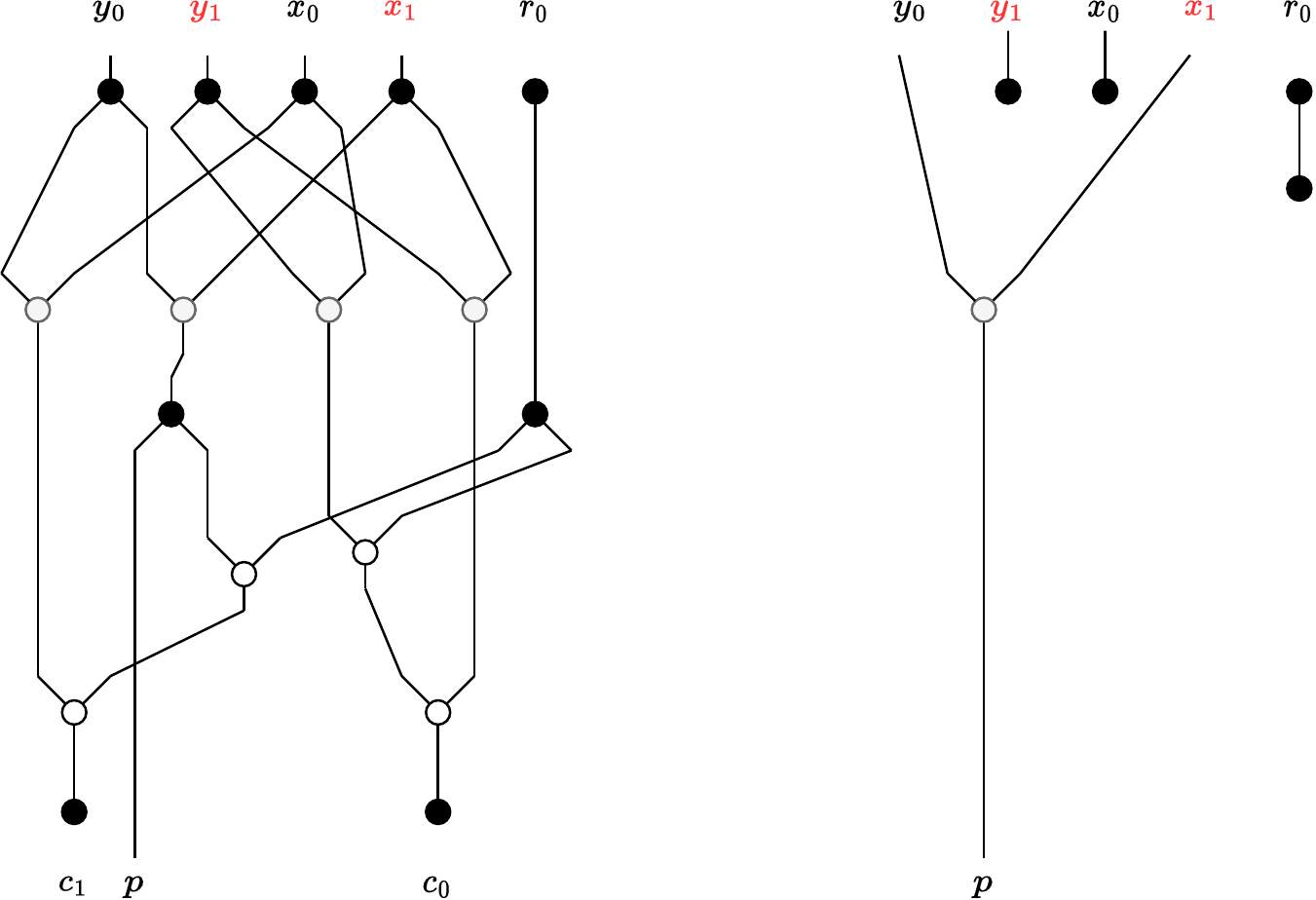}
\caption{Example of DOM with two shares per domain. The two domains are colored black and red.\label{fig:cassiers}}
\end{figure}

The example in Figure \ref{fig:cassiers} is taken from \cite{cassiersTriviallyEfficientlyComposing2020}. It shows that DOM multiplication is not
PINI because there is one probe $p$ should be allowed to block 
only one domain from being reached by an erase morphism, instead of two.

\hypertarget{robust-probing-security}{%
\subsection{Robust probing security}\label{robust-probing-security}}

Concerning the conventional (overly conservative) definition of glitch-robust probing security, we could extend the prop \finprobr{} with an additional morphism which corresponds to a register and impose additional equations which, for example, allow the cut rule only in presence of the register itself. 
However, in this timeless description the correlation matrix of a register is the identity; thus the string diagram tool is no longer sound with respect to the probability distributions. Two string diagrams would, in fact, correspond to the same correlation matrix, but they would not be reducible to one another.

\hypertarget{conclusions}{%
\section*{Conclusions}\label{conclusions}}
\addcontentsline{toc}{section}{Conclusions}

In conclusion, this paper has provided a formalization of simulatability in category-theoretic terms, which could potentially lead to better tools in the field of side-channel attacks, countermeasure analysis, and probing security. I believe that these findings are significant and could open up new forms of reasoning in the field. However, practical implementations of the proposed tools are needed to fully understand the implications of this work. Therefore, I call on all willing researchers to collaborate with me and develop these new tools in order to further explore the potential of these ideas.

\bibliographystyle{plain}
\bibliography{biblio}

\end{document}